\documentclass[useAMS,usenatbib,letters]{mn2e}

\usepackage{graphicx} 
\usepackage{txfonts}
\usepackage{lscape}
\usepackage{mathrsfs}
\usepackage{nicefrac}

\title[Lensing mass estimate of a damped Ly-$\alpha$ galaxy at z=2.2]{First gravitational lensing mass estimate of a damped Lyman-$\alpha$ galaxy at $z=2.2$}

\author[C. Grillo and J. P. U. Fynbo]{C. Grillo$^{1}$\thanks{E-mail: grillo@dark-cosmology.dk} and J. P. U. Fynbo$^{1}$ \\
$^{1}$Dark Cosmology Centre, Niels Bohr Institute, University of Copenhagen, Juliane Maries Vej 30, 2100 Copenhagen, Denmark}

\begin{document}

\date{Accepted. Received; in original form}

\pagerange{\pageref{firstpage}--\pageref{lastpage}} \pubyear{}
    
\maketitle

\label{firstpage}

\begin{abstract}

We present the first lensing total mass estimate of a galaxy, at redshift 2.207, that acts as a gravitational deflector and damped Lyman-$\alpha$ absorber on the background QSO SDSS J1135$-$0010, at redshift 2.888. The remarkably small projected distance, or impact parameter, between the lens and the source has been estimated to be $0.8\pm0.1$ kpc in a recent work. By exploiting the Sloan Digital Sky Survey database, we establish a likely lensing magnification signal in the photometry of the QSO. This is determined to be 2.2 mag brighter (or 8 times more luminous) than the median QSO at comparable redshifts. We describe the total mass distribution of the lens galaxy with a one-component singular isothermal sphere model and contrast the values of the observed and model-predicted magnification factors. For the former, we use conservatively the photometric data of the 95\% of the available distant QSO population. We estimate that the values of the lens effective velocity dispersion and two-dimensional total mass, projected within a cylinder with radius equal to the impact parameter, are included between 60 and 170 km~s$^{-1}$ and 2.1$\times 10^{9}$ and 1.8$\times 10^{10}$ M$_{\odot}$, respectively. We conclude by remarking that analyses of this kind are crucial to exploring the relation between the luminous and dark matter components of galaxies in the high-redshift Universe.

\end{abstract}


\section{Introduction}

A fundamental goal in modern astrophysics is a thorough understanding of the evolution over cosmic history of the physical properties of galaxies. Any successful theory of structure formation should be able to accommodate the observational results on the evolution of the galaxy stellar mass assembly, chemical composition, and relation between luminous and dark-matter components. At cosmologically relevant redshifts ($z \gtrsim 2$), studies of Damped Lyman-$\alpha$ Absorbers (DLAs) constitute a very interesting empirical basis for testing theories of structure formation. DLAs are defined as quasar (QSO) absorbing systems with H {\small I} column densities $N_{\mathrm{H\,I}} \geq 2\times 10^{20}$ cm$^{-2}$. At such column densities, the Ly$\alpha$ absorption optical depth is so large that the QSO continuum is entirely absorbed at the redshifted Ly$\alpha$ line wavelength and the line profile is dominated by the damping wings due to Lorentz broadening. At redshifts $z \gtrsim 2$, DLAs dominate the neutral hydrogen budget in the Universe, thus representing a major source for star formation at those redshifts. Their neutral gas content can account for a significant fraction of the stellar mass observed in local galaxies. Hence, being considered important building blocks of present day galaxies, DLAs are expected to trace the history of chemical evolution of galaxies (for a complete review on DLAs, see \citealt{wol05}). 

Cosmological simulations predict that the gas observed in a DLA must have cooled and collapsed and be embedded in a dark-matter halo (e.g., \citealt{gar97}; \citealt{mo98}; \citealt{nag04}; \citealt{pon08}; \citealt{rah13}). To date, we have limited and still somewhat conflicting evidence about the dark-matter haloes in which DLAs reside (e.g., \citealt{led06}; \citealt{coo06}; \citealt{mol13}). This is mainly due to the difficulty in measuring the total mass of these systems, which are often not even identified in the available optical images, because of the overwhelming light contribution of the background QSO. In a few DLAs, rough estimates of their virial mass have been obtained by combining size and velocity dispersion measurements (e.g., \citealt{kro13}; \citealt{fyn13}). The latter are usually inferred from the spectral widths of emission or absorption lines. The kinematical scenario is not easy to interpret though, since the line widths likely result from a combination of infall/outfall, random, and rotational motions. Theoretically, in some fortunate cases, gravitational lensing could offer a valid alternative for a total mass measurement, independent from equilibrium and geometrical hypotheses on the mass distribution. In fact, if sufficiently large, the lensing magnification of a background QSO by a foreground DLA could be used to estimate the DLA total mass projected within a cylinder with radius equal to the impact parameter. In practice, most of the times the DLA is not close (in projection) and/or massive enough to produce a detectable magnification effect.

\begin{figure}
  \centering
  \includegraphics[width=0.46\textwidth]{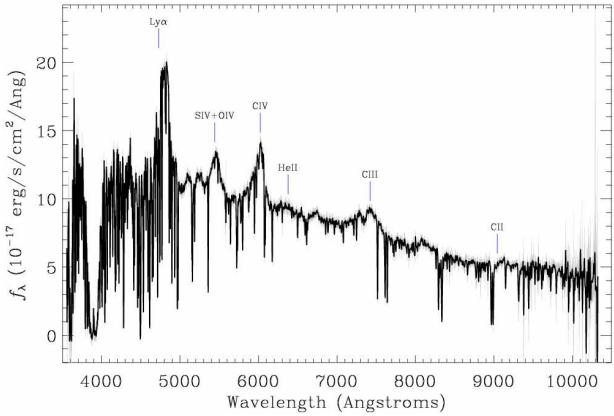}
  \caption{SDSS DR10 spectrum of the QSO J1135$-$0010 measured within a circular aperture of 3 arcsec in diameter. Several highlighted emission lines have been used to determine a value of 2.888 for the redshift of the QSO. The Ly$\alpha$ absorption line produced by the DLA galaxy at redshift 2.207 and studied in this paper is visible at $\approx$3900 \AA.}
  \label{fig1}
\end{figure}

We present here the extraordinary case of the QSO SDSS J113520.39$-$001053.5 (hereafter J1135$-$0010; see Fig. \ref{fig1} and Table~\ref{tab1}), at redshift 2.888, that is magnified by the gravitational lensing effect produced by a foreground DLA, at redshift 2.207 (\citealt{not12}; \citealt{kul12}; see also the clear absorption feature at $\approx$3900 \AA $\,$ in Fig. \ref{fig1}). A detailed study of the available data has suggested that the DLA is a young, gas-rich, compact starburst galaxy. In this system, we can obtain the first lensing magnification estimate because of the very small DLA impact parameter of $0.10 \pm 0.01$ arcsec (see \citealt{not12}) and thanks to the large QSO dataset available from the Sloan Digital Sky Survey (SDSS) Data Release Ten (DR10).

This letter is organised as follows. In Sect. 2 we describe how we estimate the lensing magnification signal. In Sect. 3, we illustrate the modelling of the total mass distribution of the DLA to predict the lensing magnification factor. In Sect. 4, we show our results on the projected total mass measurement of the DLA. In Sect. 5, we summarise our analysis and discuss the results of previous works and future prospects. Throughout this study, we adopt a standard $\Lambda$CDM cosmology ($H_{0}=70$ km s$^{-1}$ Mpc$^{-1}$, $\Omega_{m}=0.3$, and $\Omega_{\Lambda}=0.7$). In this model, 1 arcsec corresponds to a linear size of 8.26 kpc at the DLA redshift. All magnitudes are given in the AB system.

\begin{table*}
\centering
\caption{Photometric and spectroscopic properties of the QSO J1135$-$0010.}
\label{tab1}
\begin{tabular}{ccccccccc} 
\hline\hline \noalign{\smallskip}
Object & R.A. & Dec. & $z_{\mathrm{sp}}$ & $u$ & $g$ & $r$ & $i$ & $z$ \\
& (J2000) & (J2000) & & (mag) & (mag) & (mag) & (mag) & (mag) \\
\noalign{\smallskip} \hline \noalign{\smallskip}
J1135$-$0010 & 11:35:20.39 & $-$00:10:53.5 & 2.888 & $21.15 \pm 0.07$ & $18.95 \pm 0.01$ & $18.41 \pm 0.01$ & $18.20 \pm 0.01$ & $18.22 \pm 0.02$ \\
\noalign{\smallskip} \hline
\end{tabular}
\begin{list}{}{}
\item[\textbf{Note.}] Magnitudes are extinction-corrected.
\end{list}
\end{table*}

\section{Measurement}

\begin{figure}
  \centering
  \includegraphics[width=0.24\textwidth]{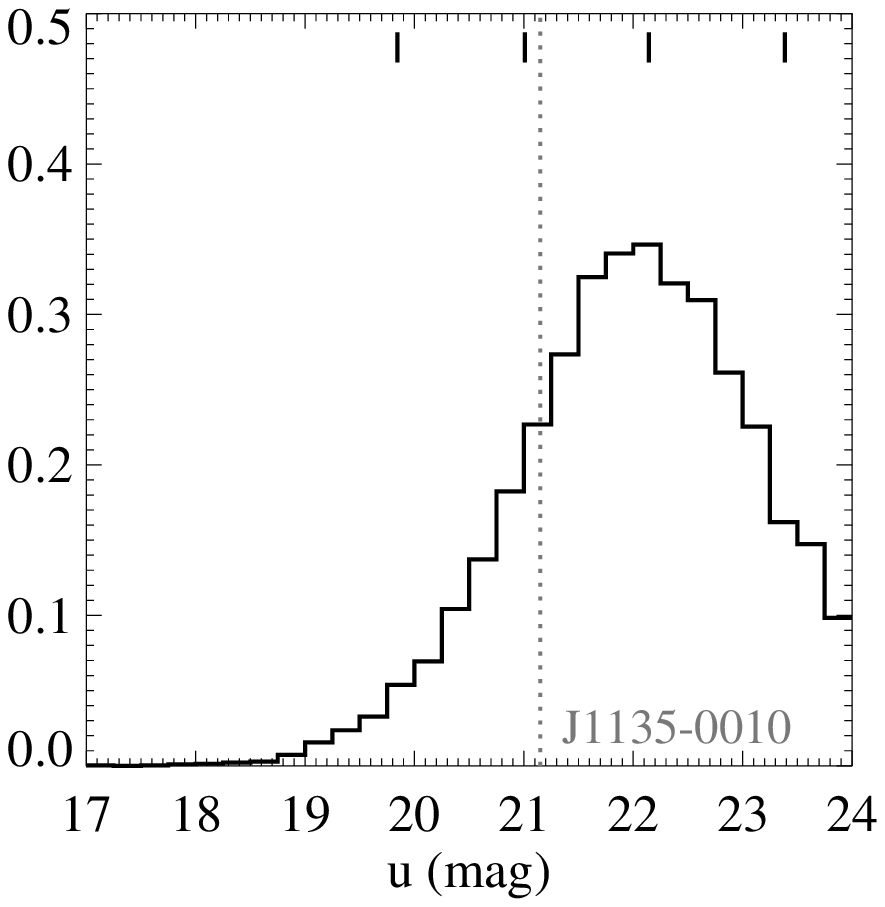}
  \includegraphics[width=0.24\textwidth]{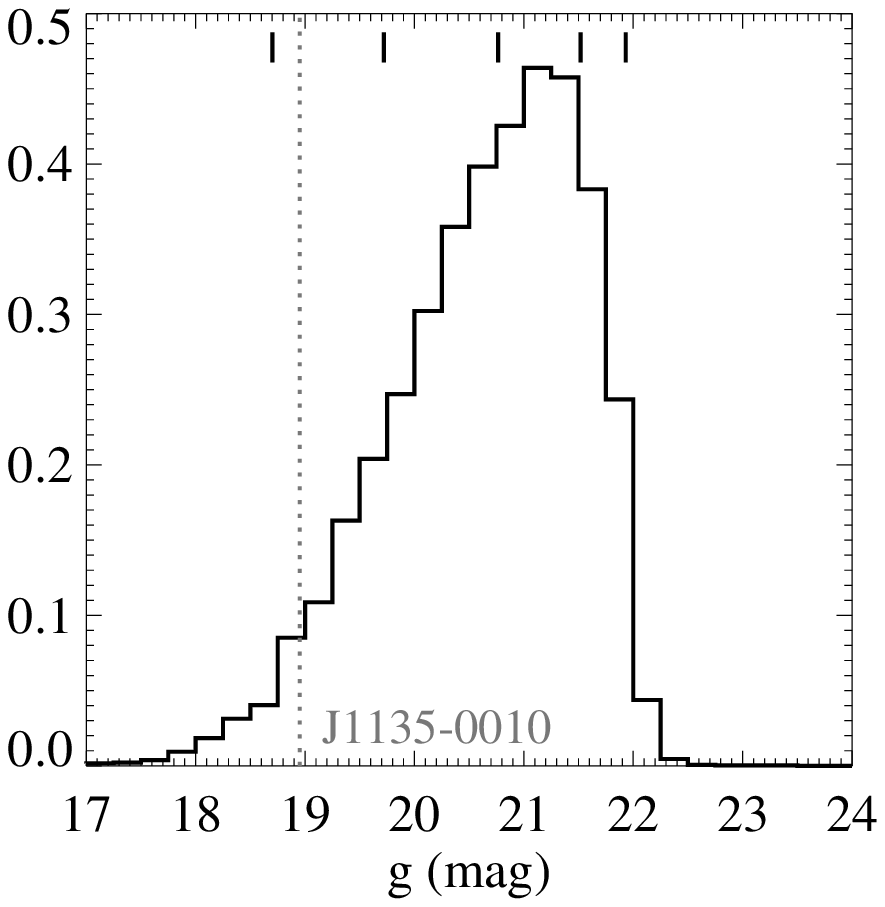}
  \includegraphics[width=0.24\textwidth]{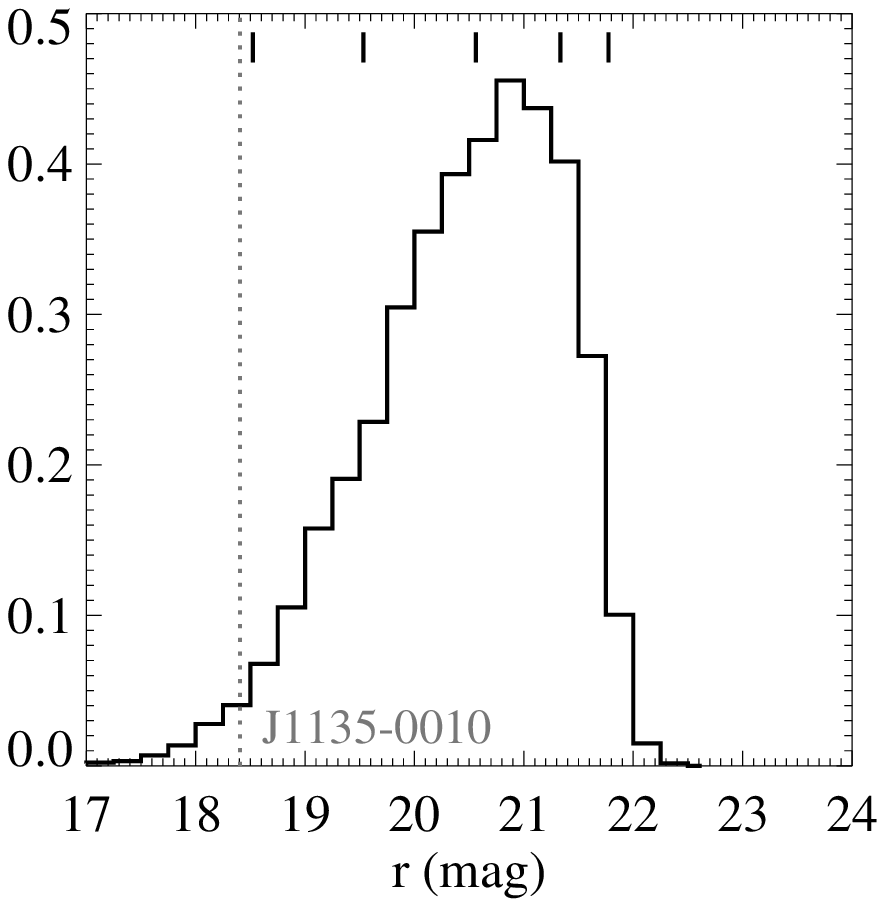}
  \includegraphics[width=0.24\textwidth]{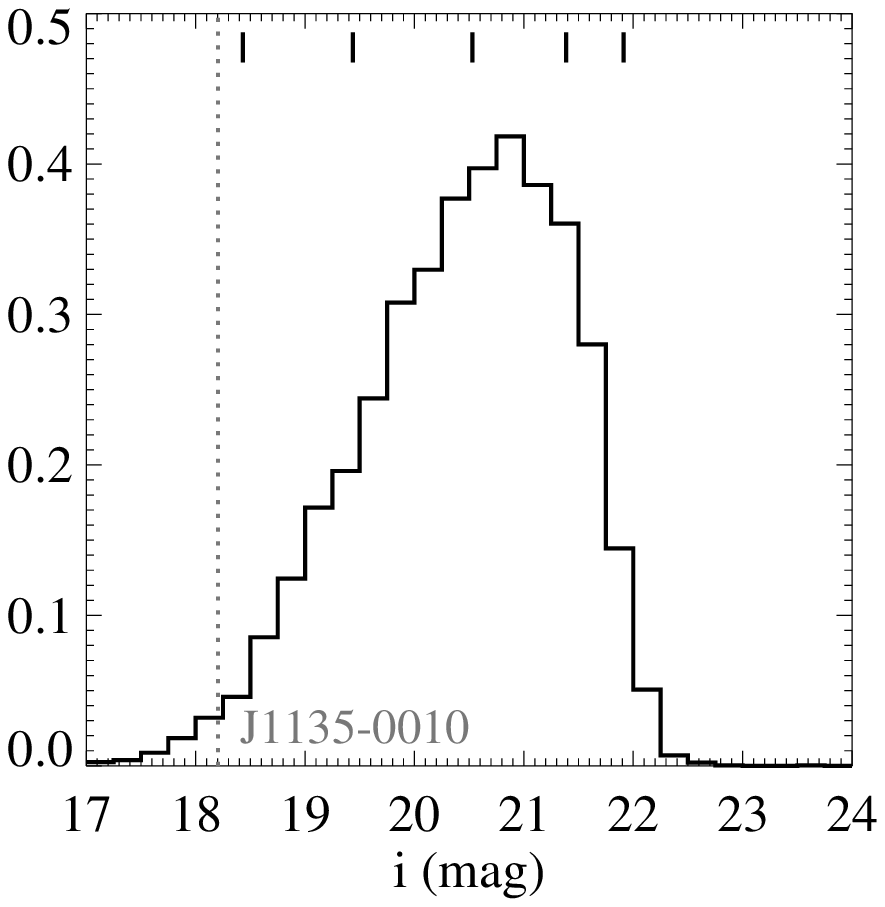}
  \includegraphics[width=0.24\textwidth]{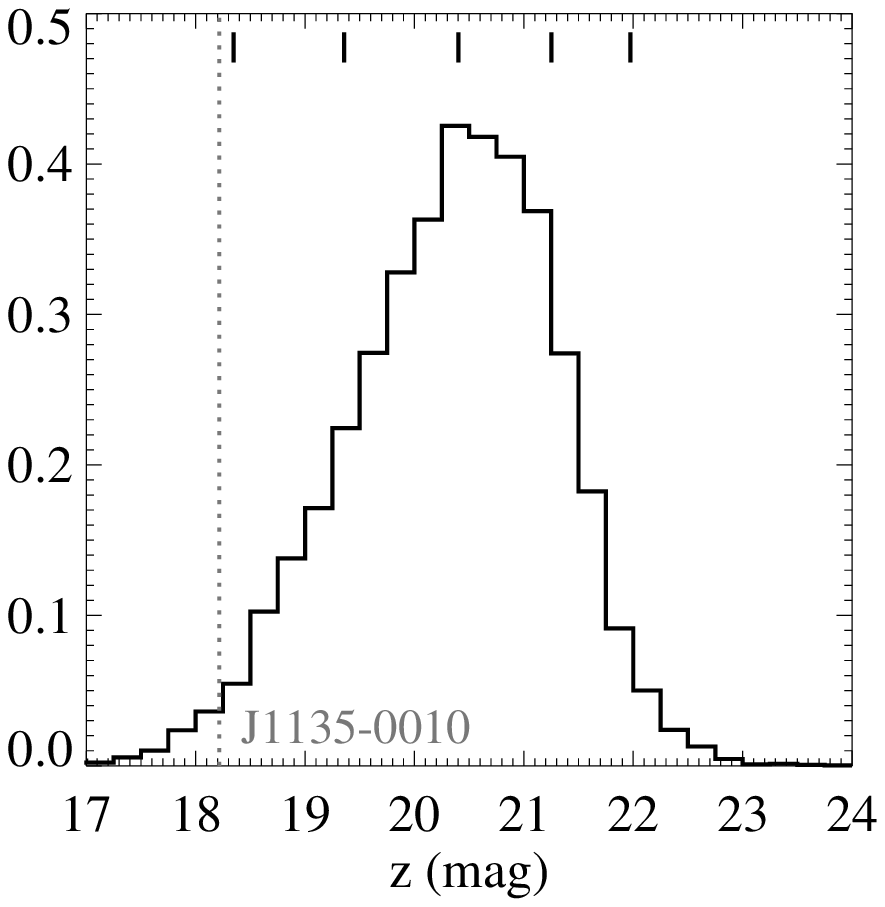}
  \caption{Probability distribution functions of the extinction-corrected $u$, $g$, $r$, $i$, and $z$ magnitudes of the SDSS DR10 sample of 11512 QSOs with spectroscopic redshifts between 2.788 and 2.988. The 5 black vertical bars on the top of each panel represent the magnitude values of, respectively, the 2.5\%, 16\%, 50\%, 84\%, and 97.5\% levels of the sorted elements in the sample. The magnitude values of the QSO J1135$-$0010 are shown with grey dotted lines.}
  \label{fig2}
\end{figure}

\begin{table*}
\centering
\caption{Values of the magnitude differences, $\Delta$, in the $u$, $g$, $r$, $i$, and $z$ SDSS bands between the sorted QSOs (at different per cent levels) and J1135$-$0010. The statistical sample consists of 11512 QSOs with spectroscopic redshifts between 2.788 and 2.988 (in parentheses, 1207 QSOs between 2.878 and 2.898).}
\label{tab2}
\begin{tabular}{ccccccc} 
\hline\hline \noalign{\smallskip}
Level & $\Delta_{u}$ & $\Delta_{g}$ & $\Delta_{r}$ & $\Delta_{i}$ & $\Delta_{z}$ & $\Delta m$ \\
& (mag) & (mag) & (mag) & (mag) & (mag) & (mag) \\
\noalign{\smallskip} \hline \noalign{\smallskip}
2.5\% & $-$1.31 ($-$1.21) & $-$0.25 ($-$0.14) & 0.12 (0.23) & 0.23 (0.33) & 0.13 (0.24) & 0.16 (0.27) \\
16\% & $-$0.14 ($-$0.15) & 0.77 (0.77) & 1.13 (1.10) & 1.23 (1.20) & 1.14 (1.14) & 1.17 (1.15) \\
50\% & 0.99 (0.95) & 1.81 (1.80) & 2.16 (2.15) & 2.33 (2.31) & 2.19 (2.21) & 2.23 (2.22) \\
84\% & 2.24 (2.19) & 2.57 (2.58) & 2.93 (2.95) & 3.18 (3.18) & 3.04 (3.04) & 3.04 (3.05) \\
97.5\% & 3.74 (3.78) & 2.98 (2.98) & 3.37 (3.38) & 3.71 (3.68) & 3.76 (3.74) & 3.53 (3.52) \\
\noalign{\smallskip} \hline
\end{tabular}
\end{table*}

We start by extracting from the SDSS DR10 database a catalogue with the values of the extinction-corrected $u$, $g$, $r$, $i$, and $z$ magnitudes for the spectroscopic sample of QSOs that lie in the redshift interval between $z_{\mathrm{sp}}-\delta z_{\mathrm{sp}}$ and $z_{\mathrm{sp}}+\delta z_{\mathrm{sp}}$, with $z_{\mathrm{sp}}$ and $\delta z_{\mathrm{sp}}$ equal to, respectively, 2.888 (see Table \ref{tab1}) and 0.1. We find 11512 objects that meet these criteria. Then, in each of the 5 bands we measure the magnitude differences ($\Delta_{u}$, $\Delta_{g}$, $\Delta_{r}$, $\Delta_{i}$, and $\Delta_{z}$) between the 2.5\%, 16\%, 50\%, 84\%, and 97.5\% levels of the sample QSOs, sorted in magnitude, and J1135$-$0010.

We show the outcomes of this analysis in Fig. \ref{fig2} and Table \ref{tab2}. In Fig. \ref{fig2}, we plot the probability density distributions of the magnitude values of the sample QSOs in the 5 SDSS photometric bands and highlight the magnitude values of J1135$-$0010. We find that approximately 19\% and 5\% of all QSOs are brighter than J1135$-$0010 in the $u$ and $g$ bands, respectively, but only less than 2\% at longer wavelengths, i.e., in the $r$, $i$, and $z$ bands. This matches very well with the picture that J1135$-$0010 is substantially magnified by the DLA galaxy at redshift 2.207. In fact, although J1135$-$0010 is relatively brighter than most QSOs in the same redshift range in the $u$ and $g$ bands, here the absorption of the DLA around 3900 \AA $\,$, visible in Fig. \ref{fig1} and affecting the magnitudes measured in these two filters, makes the lensing magnification effect less evident. In the $r$, $i$, and $z$ bands instead, the unabsorbed continuum of J1135$-$0010 is measured more accurately and its intensity appears significantly higher than that of the average QSO. More quantitatively, from Table \ref{tab2} (third row) we see that in the $r$, $i$, and $z$ filters J1135$-$0010 is approximately 2.2 mag brighter (i.e., 8 times more luminous) than the median QSO in the sample. Looking at the same table, we remark that in the 3 reddest bands J1135$-$0010 is brighter (i.e., the $\Delta$ values are positive) than the 68\% and 95\% of the sample QSOs included in the intervals measured around the median values. Furthermore, we note that in the 2 bluest bands J1135$-$0010 is fainter (i.e., the $\Delta$ values are negative) than the brightest (2.5\%) QSOs in the sample, as a consequence of the fact that the latter are almost certainly not affected by absorption features from DLAs at redshifts of about 2.2.

In Table \ref{tab2}, we also report the values of $\Delta m$, that are defined as the weighted mean values of $\Delta_{r}$, $\Delta_{i}$, and $\Delta_{z}$, where the weights are calculated from the magnitude uncertainties of J1135$-$0010 (see Table \ref{tab1}). We consider these quantities the most robust estimates, derived in the discussed statistical way, of the magnification factors of J1135$-$0010 at different per cent levels.

We check that the magnification estimates are not dependent on the extracted QSO sample. To do that, we repeat the measurements presented above starting from three additional QSO samples with $\delta z_{\mathrm{sp}}$ equal to 0.05, 0.01, and 0.005. The values shown in parentheses in Table \ref{tab2} refer to the sample obtained by reducing $\delta z_{\mathrm{sp}}$ from the previous value of 0.1 to 0.01, resulting in 1207 objects in total. From this test, we confirm that our measurements are not sensitive to the redshift selection of the SDSS QSOs.

We mention that we have also tried to estimate J1135$-$0010 intrinsic luminosity, thus the QSO lensing magnification factor, in alternative ways. We have investigated several empirical correlations between some spectral features and the luminosity values of QSOs. For example, we have looked into the so-called Baldwin effect (\citealt{bal77}; see also, e.g., \citealt{bia12}), an anti-correlation between the equivalent width of the C {\small IV} emission line and its nearby continuum luminosity, and the correlation between the velocity offset of the C {\small IV} with respect to the Mg {\small II} lines and luminosity (e.g., \citealt{ric02}). Unfortunately, all investigated QSO empirical relations between spectroscopic quantities, not affected by the lensing magnification effect, and luminosity have significant intrinsic scatter. Because of that, although we could verify that J1135$-$0010 is located in regions of the parameter space consistent with the hypothesis of high magnification, we were not able to find a method more precise than that discussed above to measure the QSO lensing magnification factor.

In conclusion, despite the fact that our lensing magnification estimate is based on the QSO photometry only, the assumption that J1135$-$0010 is a {\it normal} QSO significantly magnified by the DLA lens galaxy seems much more likely than that of an extremely luminous QSO with a negligible magnification factor. The two main reasons are the spectroscopic confirmation of the existence of a DLA lens galaxy with a very small impact parameter and the observation of just a few brighter QSOs, some of which clearly show in the SDSS images possible magnifying lens galaxies very close in projection.

\section{Modelling}

We describe the total mass distribution of the DLA galaxy in terms of a one-component singular isothermal sphere (SIS; $\rho \propto r^{-2}$) model. This profile has been shown to represent well the total mass distribution of massive early-type galaxies (e.g., \citealt{koo09}; \citealt{bar09}) and it is adopted here mainly because of its parametrisation simplicity. In fact, a SIS model is entirely characterised by the value of an effective velocity dispersion $\sigma_{\mathrm{SIS}}$. This last quantity should not be confused with the velocity dispersion of stars used in dynamical studies of galaxies. Nonetheless, several analyses of massive elliptical galaxies have demonstrated that a good estimator of $\sigma_{\mathrm{SIS}}$ is the central velocity dispersion of the stellar component (e.g., \citealt{tre06}; \citealt{gri08}). The three-dimensional (total) density profile, $\rho_{\mathrm{T}}(r)$, of a SIS model is given by
\begin{equation}
\rho_{\mathrm{T}}(r) = \frac{\sigma^{2}_{\mathrm{SIS}}}{2 \pi G r^{2}} \, ,
\label{eq:01}
\end{equation}
where $G$ is the gravitational constant. By projecting $\rho_{\mathrm{T}}(r)$ along the line of sight and integrating on the plane perpendicular to that direction, one finds that the total projected (two-dimensional) mass enclosed within a cylinder of radius $\tilde{R}$ is
\begin{equation}
M_{\mathrm{T}}(R<\tilde{R}) = \frac{\pi\sigma^{2}_{\mathrm{SIS}}\tilde{R}}{G} \, .
\label{eq:02}
\end{equation}

From the gravitational lensing theory, it can be shown that for a SIS model the value of the lens Einstein angle $\theta_{\mathrm{Ein}}$ (i.e., the radius of the circle, on the lens plane, into which a source perfectly aligned with the lens and observer is imaged) is related to that of $\sigma_{\mathrm{SIS}}$ in the following way:
\begin{equation}
\theta_{\mathrm{Ein}} = 4 \pi\, \Big( \frac{\sigma_{\mathrm{SIS}}}{c} \Big)^2 \frac{D_{ls}}{D_{os}} \, ,
\label{eq:03}
\end{equation}
where $c$ is the speed of light and $D_{ls}$ and $D_{os}$ are, respectively, the angular diameter distances between the lens and the source and the observer and the source. Moreover, a background source that is observed as a single lensed image (i.e., not in the strong lensing regime, where multiple images of one source are observable) at an angular distance of $\theta$ larger than $\theta_{\mathrm{Ein}}$ from the SIS lens centre is magnified by a factor $\mu(\theta)$ equal to
\begin{equation}
\mu(\theta) = 1 + \frac{\theta_{\mathrm{Ein}}}{\theta-\theta_{\mathrm{Ein}}} .
\label{eq:04}
\end{equation}
This means that in the absence of the gravitational lensing effect a source would be observed with an intrinsic (\emph{unlensed}) magnitude $m_{\mathrm{u}}$, but when it is lensed by a deflector located at a projected angular distance $\theta$, the same source is magnified and observed with a (\emph{lensed}) magnitude $m_{\mathrm{l}}(\theta)$ that is
\begin{equation}
m_{\mathrm{l}}(\theta) = m_{\mathrm{u}} - 2.5 \times \log\big[\mu(\theta)\big]  .
\label{eq:05}
\end{equation}
In summary, the observed magnitude of a lensed source will depend on the geometrical lensing configuration (i.e., on the redshifts of lens and source and angular distance between lens and source) and on the total mass of the lens, in addition to the source intrinsic magnitude.

\section{Results}

\begin{figure*}
  \centering
  \includegraphics[width=0.37\textwidth]{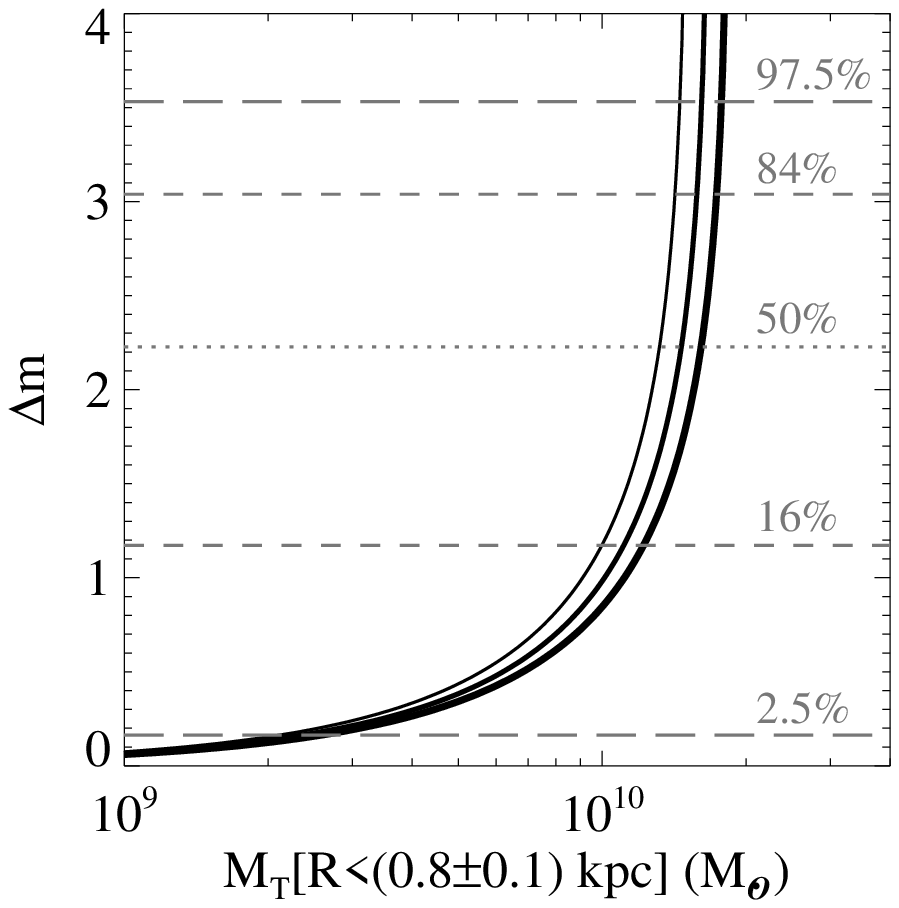}
  \includegraphics[width=0.37\textwidth]{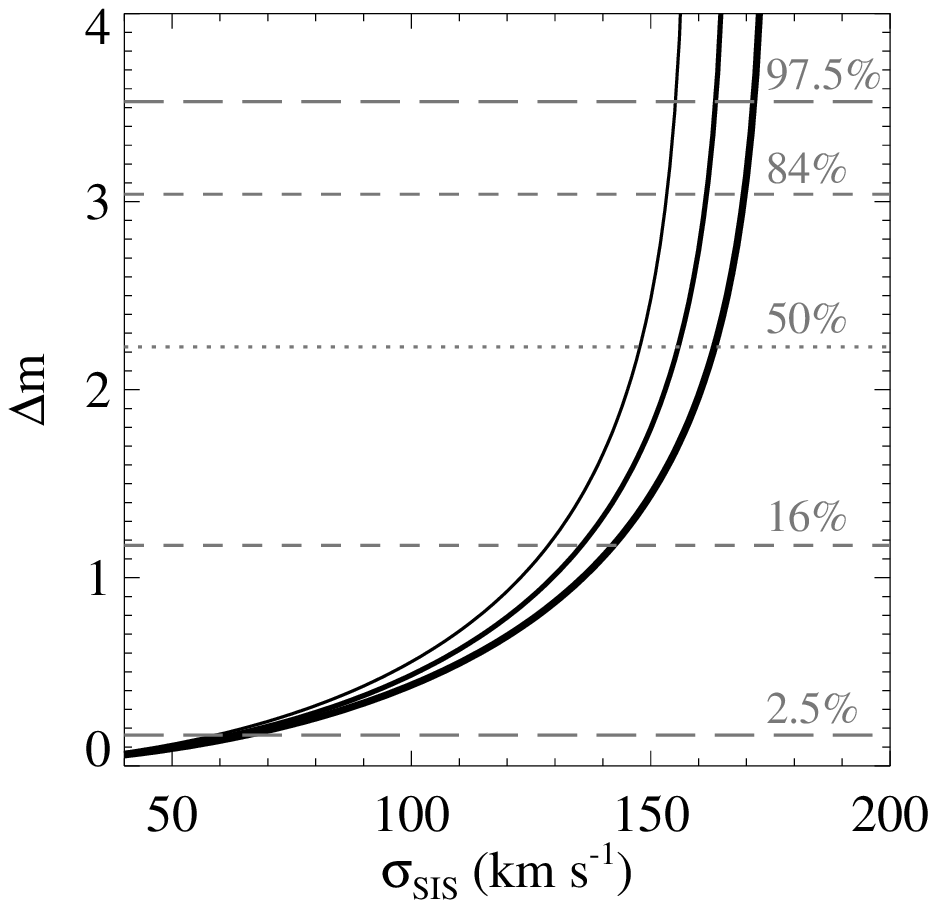}
  \caption{Values of model-predicted and observed magnitude difference (i.e., magnification) for the DLA galaxy lensing the QSO J1135$-$0010. The three increasingly thicker black curves show, respectively, the expected magnification factors for a SIS model located at projected angular distances from the observed image of 0.09, 0.10, and 0.11 arcsec (i.e., 0.7, 0.8, and 0.9 kpc at the DLA redshift of 2.207) as a function of the lens total projected (two-dimensional) mass within these radii (\emph{on the left}) and effective velocity dispersion (\emph{on the right}). The grey lines show the estimated magnification factors, at different per cent levels, from a sample of 11512 SDSS QSOs with spectroscopic redshifts between 2.788 and 2.988 (see Table \ref{tab2}).}
  \label{fig3}
\end{figure*}

By comparing the model-predicted values of the lensing magnification (i.e., $m_{\mathrm{u}}-m_{\mathrm{l}}(\theta)=2.5 \times \log\big[\mu(\theta)\big]$, see Eq. (\ref{eq:05})) and the estimated values of the magnification in J1135$-$0010 (see Sect. 2), we measure the total mass of the DLA galaxy at redshift 2.207 projected within a cylinder with radius given by the estimated impact parameter $b$ of $0.10 \pm 0.01$ arcsec (i.e., $0.8\pm0.1$ kpc at the DLA redshift). In Fig. \ref{fig3}, the three curves show the values of $m_{\mathrm{u}}-m_{\mathrm{l}}(b)$ as a function of total projected mass $M_{\mathrm{T}}$, or equivalently effective velocity dispersion $\sigma_{\mathrm{SIS}}$, for values of $b$ of 0.09, 0.10 and 0.11 arcsec. In the same figure, we also plot the values of $\Delta m$ from Table 2 at the different per cent levels. Within the adopted SIS mass model and using the values of $\Delta m$ from the 95\% interval, we estimate that the DLA has a total mass projected inside $0.8 \pm 0.1$ kpc extending between 2.1$\times 10^{9}$ and 1.8$\times 10^{10}$ M$_{\odot}$ and a corresponding effective velocity dispersion ranging from 60 to 170 km s$^{-1}$.

We note that 170 km s$^{-1}$ is a real upper limit for the value of the lens effective velocity dispersion. In fact, values larger than that would result in Einstein radii larger than the impact parameter and would, as a result, predict the existence of a second, multiple image of the QSO on the other side of the DLA mass centre. Given the available spectroscopic data, we can only detect one image of the QSO. 

Our measurement of the total mass of the DLA could be particularly useful for studies on the evolution of the luminous over total mass fraction in galaxies (e.g., \citealt{gri09}; \citealt{gri10}), if the DLA half-light radius were confirmed to be comparable to 0.8 kpc.

We have checked that our conservative total mass estimates are consistent with those obtained by using a point-like instead of a SIS one-component mass model for the DLA. We note that more realistic, two-component mass models (for instance with a disk and a halo components) could predict fairly different magnification factors at the same projected distance between the lens and the source, and therefore result in different total mass estimates for the DLA (e.g., \citealt{bar96}; \citealt{mal97}; \citealt{sme97}). Unfortunately, these models cannot be explored having at disposal only one physical observable, i.e. the magnification value of a source that is not multiply lensed.

Finally, we caution that J1135$-$0010 could be a particularly rare and intrinsically highly-luminous QSO or in an extremely active phase. If this were the case, our magnification measurements would be overestimated and with them also the projected total mass of the DLA. However, we remark that non-extreme cases can be accommodated by our conservative mass intervals.

\section{Discussion and conclusions}

The main results of this work can be summarized as follows:

\begin{itemize}

\item[$\bullet$] Starting from a sample of 11512 QSOs in the SDSS DR10 database, at redshifts similar to that (2.888) of J1135$-$0010 , we have measured that the latter is 2.2 mag (i.e., 8 times) brighter than the median QSO in the sample. This supports the hypothesis that J1135$-$0010 is gravitationally magnified in a significant way by the DLA galaxy (at redshift 2.207) that was previously identified by inspecting the QSO spectrum.

\item[$\bullet$] A one-component SIS model for the total mass distribution of the lens DLA galaxy predicts magnification factors that are consistent with the observed ones (based on the interval of the 95\% of the sample QSOs centered on the median) if the lens total mass, projected within a cylinder with radius of $0.8\pm0.1$ kpc, and effective velocity dispersion values range, respectively, from 2.1$\times 10^{9}$ to 1.8$\times 10^{10}$ M$_{\odot}$ and from 60 to 170 km s$^{-1}$. 

\end{itemize}

Compared to the results of the analysis by \citet{not12}, our study favours a slightly more massive DLA galaxy. In the cited paper, the value of the galaxy velocity dispersion, estimated from the spectral widths of $[$O {\small III}$]$ and H$\alpha$ emission lines, is found to be approximately 50 km~s$^{-1}$ (although the metal absorption lines point to somewhat larger values), and those of the virial and neutral gas masses of the order of $10^{10}$ and a few $10^{9}$ M$_{\odot}$, respectively. This little discrepancy between the lensing and dynamical total mass measurements might reveal a small bias in one (or both) of the methods. More investigations of this kind would be needed to address properly this problem.

We conclude by remarking that gravitational lensing studies like that presented here can help obtain an accurate characterisation of the dark-matter haloes hosting DLAs. This is key to understanding the formation and evolution history of galaxies, and therefore distinguishing among different hierarchical models.

\section*{Acknowledgments}

We thank M. Vestergaard for useful discussions. The Dark Cosmology Centre is funded by the DNRF. Funding for SDSS-III has been provided by the Alfred P. Sloan Foundation, the Participating Institutions, the National Science Foundation, and the U.S. Department of Energy Office of Science. The SDSS-III web site is http://www.sdss3.org/. SDSS-III is managed by the Astrophysical Research Consortium for the Participating Institutions of the SDSS-III Collaboration including the University of Arizona, the Brazilian Participation Group, Brookhaven National Laboratory, Carnegie Mellon University, University of Florida, the French Participation Group, the German Participation Group, Harvard University, the Instituto de Astrofisica de Canarias, the Michigan State/Notre Dame/JINA Participation Group, Johns Hopkins University, Lawrence Berkeley National Laboratory, Max Planck Institute for Astrophysics, Max Planck Institute for Extraterrestrial Physics, New Mexico State University, New York University, Ohio State University, Pennsylvania State University, University of Portsmouth, Princeton University, the Spanish Participation Group, University of Tokyo, University of Utah, Vanderbilt University, University of Virginia, University of Washington, and Yale University.


\label{lastpage}

\end{document}